\documentclass[12pt]{article}
\usepackage{graphicx,amsmath,amssymb,cite}
\usepackage{epsfig,epsf}
\usepackage{graphicx}
\usepackage{epstopdf}
\newcommand{\beq}{\begin{equation}}
\newcommand{\eeq}{\end{equation}}

\numberwithin{equation}{section}

\usepackage{epsfig}
\usepackage{cite}
\usepackage{color}

\newcommand{\be}{\begin{equation}}
\newcommand{\ee}{\end{equation}}
\newcommand{\beeq}{\begin{eqnarray}}
\newcommand{\eeeq}{\end{eqnarray}}

\def\bea{\begin{eqnarray}}
\def\eea{\end{eqnarray}}

\usepackage{textcomp}

\begin{document}

\begin{center}
~~~~~~~~~~~~~~~~~~~~~~~~~~~~~~~~~~~~~~~~~~~~~~~~~~~~~~~~~~~~~~~~~~~~~~~~~~~~~~~~~~~DESY Report 16-231\\
\vskip 1.0cm
{\Large{\bf Dipole model analysis of highest precision HERA data,  including very low $Q^2$'s  }}


{\large A. Luszczak$^1$, H. Kowalski~$^2$ }\\ [0.5cm]

{\it $1$ T.Kosciuszko Cracow University of Technology, 30-067 Cracow, Poland}\\[0.1cm] 
{\it $2$ Deutsches Elektronen-Synchrotron DESY, D-22607 Hamburg, Germany}\\[0.1cm] 
 \end{center}

\vspace*{3 cm}

\begin{abstract}
 We analyse, within a dipole model, the final, inclusive  HERA DIS  cross section data in the low $x$ region, using fully correlated errors. We show, that these highest precision data are very well described within the dipole model framework starting from  $Q^2$ values of  3.5  GeV$^2$  to the highest values of $Q^2 =$  250 GeV$^2$.
 
 To analyze the saturation effects we evaluated the data including also the  very low $ 0.35 <Q^2 $   GeV$^2$  region. The fits including this region show a preference of the saturation ansatz.  
   \end{abstract}


\section{Introduction}

In our last paper~\cite{kowlusz} we investigated the high precision, inclusive HERA data~\cite{H1ZEUS} within the dipole model. In the present paper we continue this investigation using the highest precision data~\cite{hiH1ZEUS}, evaluated with the {\it correlated} errors.

Many investigations have shown that  HERA inclusive and diffractive DIS cross sections are very well described by the dipole models~\cite{BGK,GBW,Iancu:2003ge}, which provide a natural description of QCD reaction in the  low-$x$  and low $Q^2$ region. They allow a simultaneous description of many different physics reactions, like inclusive DIS processes, inclusive diffractive processes, exclusive $J/\psi$, $\rho, \phi$ production, diffractive jet production, or diffractive and non-diffractive charm production. Due to the optical theorem, all these processes are determined by the same, universal, gluon density~\cite{MSM,KT,KMW}.  The understanding of the properties of the gluon density and its precise knowledge is very important because the QCD-evolved gluon density determines the cross sections of  most relevant   physics processes, e.g. Higgs production at LHC. Any significant deviation of the predicted cross section from their Standard Model value could be a sign of new physics. 

The precise determination of the gluon density relies on the analysis of high quality, inclusive, DIS data taken over the full, experimentally accessible $x$ and $Q^2$ region.    
Therefore, the H1 and ZEUS experiments have combined their inclusive DIS cross sections which, due to a substantial reduction of systematic measurements errors, led to an increase of precision by about a factor two~\cite{H1ZEUS}. More recently, also the final set of HERA data was released which provides a full set of correlated errors. These errors contain the most complete information about the data and lead to very restrictive fits~\cite{hiH1ZEUS}. The full information about the data is made now accessible within the xFitter facility~\cite{xFitter}.

The aim of this paper is to investigate the additional information  contained in the final HERA data.  The most precise data where obtained in the region of higher Q$^2$'s (Q$^2$ from 3.5 to O(10000) GeV$^2$),  where the DGLAP evolution is known to describe  data very well. In the low $x$ region, which is investigated here, the highest achievable $Q^2$ is around 250~GeV$^2$. The investigation is performed using the so called BGK  model, see below, which uses the DGLAP evolution in the dipole scheme. The dipole approach allows to extend the perturbative description to the region of much smaller $Q^2$'s,  $0.35 < Q^2 < 3.5$ GeV$^2$. The simultaneous evaluation of the very precise data at higher $Q^2$ together with the low $Q^2$  data allows to address again the question of  high density gluonic states.

The paper is organized as follows: in Section 2 we  recall the main properties of the dipole approach and review the dipole models. In Section 3 we present the results of the  dipole and pdf fits in the higher $3.5<Q^2 < 250$ GeV$^2$ region. In Section 4 we discuss  saturation effects  including data in the lower $Q^2$ region.  In Section 5 we   summarize the results.

\section{Dipole models}
The dipole picture was first derived, in the low $x$ limit of QCD, by Nikolaev and Zaharov ~\cite{NNZ:91}. They have shown that  the deep inelastic scattering can be viewed as a two stage process; first the virtual photon fluctuates into a dipole, which consists of  a quark-antiquark pair (or a $q\bar q g$ or $q\bar q gg$ ... system) and in the second stage the dipole interacts with the proton. 
Dipole denotes a quasi-stable quantum mechanical state, which has a very long life time ($\approx 1/m_p x\;$) and a size $r$, which remains  unchanged during scattering. The wave function $\Psi$ determines the probability to find a dipole of size $r$ within a photon. This probability depends on the value of external $Q^2$ and the fraction of the photon momentum carried by the quarks forming the dipole, $z$. Neglecting the $z$ dependence, in a very rough approximathion, $Q^2 \sim 1/r^2$.

  The scattering amplitude is a product of the virtual photon wave function, $\Psi$, with the dipole cross section, $\sigma_{\text{dip}}$, which determines a probability of the dipole-proton scattering. 
Thus, within the dipole formulation of the $\gamma^* p$ scattering     
\begin{equation}
\label{edipole}
   \sigma_{T,L}^{\gamma^* p}(x,Q^{2}) = \int dr^2 \int dz \Psi^*_{T,L} (Q,r,z) \sigma_{\text{dip}}(x,r) \Psi_{T,L}(Q,r,z),
\end{equation}
where $T,L$ denotes the virtual photon polarization and $\sigma_{T,L}^{\gamma^* p}$ the total inclusive DIS cross section.

This simple and intuitive approach became then a basis of many dipole many models,
~\cite{Nemchik1996,Gotsman1995,Dosch1996,CS,Forshaw2003,
Frankfurt2005,KLMV}. 
which have been developed to test various aspects of data.  They vary due to different assumption made about the physical behavior of  dipole cross sections.  In the following we will shortly review some them to motivate the choice of the model used for present investigation.   

\subsection{GBW model}

The dipole model became an important tool in investigations of deep-inelastic scattering due to the initial observation of Golec-Biernat and W\" uesthoff (GBW) \cite{GBW}, that a simple ansatz for the dipole cross section was able to describe simultaneously the total inclusive and diffractive cross sections.

In the GBW model the dipole-proton cross section $\sigma_{\text{dip}}$ is given by
\begin{equation}
\label{eGBW}
   \sigma_{\text{dip}}(x,r^{2}) = \sigma_{0} \left(1 - \exp \left[-\frac{r^{2}}{4R_{0}^{2}(x)} \right]\right),
\end{equation}
where $r$ corresponds to the transverse separation between the quark and the antiquark, and $R_{0}^{2}$ is 
an $x$ dependent scale parameter which has a meaning of saturation radius,  $R_{0}^{2}(x)=\left(x/x_{0}\right)^{\lambda_{GBW}}/GeV^{-2}$.
The free fitted parameters are: the cross-section normalisation, $\sigma_{0}$, as well as $x_{0}$ and $\lambda_{GBW}$.  In this model saturation is taken into account in the eikonal approximation and the saturation radius is intimately related to the gluon density, see below. The exponent $\lambda_{GBW} $ determines  the growth of the total and diffractive cross section with decreasing $x$. For dipole sizes which are large in comparison to the saturation radius, $R_0$, the dipole cross section saturates by approaching a constant value $\sigma_0$, i.e. saturation damps the growth of the gluon density at low $x$. 

The GBW model provided a good description of data from medium $Q^2$ values ($\approx 30$ GeV$^2$) down to low $Q^2$ ($\approx 0.1$) GeV$^2$).
Despite its success and its appealing  simplicity  the model has some  shortcomings; in particular it describes the QCD evolution by a simple $x$ dependence,  $ \sim (1/x)^\lambda_{BGW}$, i.e the $Q^2$ dependence of the cross section evolution is solely induced by the saturation effects. Therefore, it does not  match with DGLAP QCD evolution, which is known to describe data very well from $Q^2 \approx 4$ GeV$^2$ to very large $Q^2 \approx 10000$ GeV$^2$.

\subsection{BGK model}
The evolution ansatz of the GBW model was improved in the model proposed by Bartels, Golec-Biernat and Kowalski, (BGK)~\cite{BGK},  by taking into account the  DGLAP evolution of the gluon density in an explicit way. The model preserves the GBW eikonal approximation to saturation and thus the dipole cross section is given by
\begin{equation}
\label{eBGK}
   \sigma_{\text{dip}}(x,r^{2}) = \sigma_{0} \left(1 - \exp \left[-\frac{\pi^{2} r^{2} \alpha_{s}(\mu^{2}) xg(x,\mu^{2})}{3 \sigma_{0}} \right]\right).
\end{equation}
The evolution scale $\mu^{2}$ is connected to the size of the dipole by $\mu^{2} = C/r^{2}+\mu^{2}_{0}$. This assumption allows to treat  consistently the contributions of large dipoles  without making the strong coupling constant, $\alpha_s(\mu^2$),  un-physically large. This means also that we can extend the model, keeping its perturbative character, to the data at low $Q^2$, because the external $Q^2$ and the internal $\mu^2$ scales are connected only by the wave function.  
  
 The gluon density, which  is parametrized  at the starting scale $\mu_{0}^{2}$, 
is evolved to larger scales, $\mu^2$, using LO or NLO DGLAP evolution.
 We consider here two forms of  the gluon density:
\begin{itemize}
\item 
the {\it soft} ansatz, as used in the original BGK model 
\begin{equation}
   xg(x,\mu^{2}_{0}) = A_{g} x^{-\lambda_{g}}(1-x)^{C_{g}},
\label{gden-soft}
\end{equation}
\item
the {\it soft + hard} ansatz
\begin{equation}
   xg(x,\mu^{2}_{0}) = A_{g} x^{-\lambda_{g}}(1-x)^{C_{g}}(1+D_g x +E_gx^2),
      \label{gden-softhard}
\end{equation}
\end{itemize}

The free parameters for this model are $\sigma_{0}$ and the  parameters for gluon $A_{g}$, $\lambda_{g}$, $C_{g}$ or additionally $D_g, E_g,$  
 Their values are obtained by a fit to the data. The fit results were found to be independent on the parameter  $C$, which was therefore fixed as  $C=4$ GeV$^2$, in agreement with the original BGK fits. It is also possible to vary the parameter $\mu^{2}_{0}$. However,
 to assure that the evolution is  performed in the perturbative region and to be compatible with the standard pdf fits we took  as a starting scale  $\mu^2_0 = 1.9$ or 1.1 GeV$^2$.  In the BGK model, the  $\mu_0^2$ scale is the same as the  $Q_0^2$ scale of the standard QCD pdf fits.

\section{Results of fits  in the higher $Q^2$ region}

This paper concentrates  on  the  {\it inclusive} DIS measurements in the low $x$ region, $x < 0.01$.  
Here, the contribution of the valence quarks is small, below 7\%, and has therefore been neglected for a long time. 
However now, the combined H1 and ZEUS HERA data achieve  a precision of about 2\%.  Theoretically, it is very difficult to treat valence quarks inside the dipole framework because,   the dipole amplitudes are not well defined in the region of high~$x$. 
In our previous paper~\cite{kowlusz}, we developed therefore an heuristic approach in which we  added the valence quark contribution from the standard pdf's fits to the dipole predictions. Hence, the dipole contribution plays just a role of the sea quarks in the standard pdf's. This procedure is justified by the fact that the sea quark contribution disappears at larger $x$.    

 For the purpose of this investigation we choose the BGK model, because it is expected to provide the best description of data in the higher 
 $Q^2$  range, as it uses  DGLAP evolution.  The fits were performed within the xFitter system, 
 where the dipole model and the valence quarks contributions are a part of the same framework~\cite{xFitter}. 
Therefore,  the QCD evolution is the same as in the standard xFitter pdf fits.  For gluon density we used both the  {\it soft } and  {\it soft+hard} ansatz  with the NLO evolution.

 We used for fits a complete   set of reduced cross section points,  $\sigma_r$, obtained at different electron and proton energies: HERA1+2-NCep-460, HERA1+2-NCep-575,   HERA1+2-NCep-820, HERA1+2-NCep-920 and HERA1+2-NCem.

\subsection{Fits to the  dipole model.}

The results of the BGK fit  with valence quarks and the {\it soft} gluon density, are shown in  Table~\ref{tabl1}, \ref{tabl2} and  \ref{tabl3}. 
The best fit is obtained  using valence quarks and gluon density  of the {\it soft + hard} type,  see Table~\ref{tabl7} and~\ref{tabl9}. In Table~\ref{tabl1},~\ref{tabl2} and~\ref{tabl7} the starting QCD scale is $Q_0^2 = 1.9$ GeV$^2$, in Table~\ref{tabl9} it is   $Q_0^2 = 1.1$ GeV$^2$,

In Table~\ref{tabl1},~\ref{tabl7} and~\ref{tabl9}  the valence quark contribution was taken from the standard pdf fit, shown below, in Table~\ref{tabl2} the valence quarks were fitted. 
In these tables, $N_{df}$ denotes the number of degrees of freedom, which is equal to the number of measured data points  minus the number of free parameters used in the fit. The parameters  
 $\sigma_{0}$ of the dipole model and the starting parameters for gluon $A_{g}$, $\lambda_{g}$, $C_{g}$ are obtained from the fit. The value of the parameter $C$ was fixed, as explained above.  
 To limit the fit to the perturbative region only  we took $Q^2 \ge Q^2_{min}$ with $Q^2_{min} = 3.5$ (or 8.5) GeV$^2$. For the $x$ region, we took $x\le 0.01$. There are 538 (or 452) measured points in this region. 

\begin{table}[htbt]
\begin{center}
\begin{tabular}{|c|c|c|c|c|c|c|c|c|c|c|} 
\hline 
 $Q_{min}^2$  [GeV$^2$]&
$\sigma_0[mb]$ & $A_g$ & $\lambda_g$ & $C_g$ & $N_{df}$& $\chi^2$& $\chi^2/N_{df}$\\
\hline
 $3.5 $&  87.0$\pm$ &  2.32$\pm$ & -0.056$\pm$ & 8.21$\pm$&  534 &  551.1 &  1.03 \\
  & 8.9 & 0.009&  0.11&  0.80 & & &  \\
   8.5 &  72.4$\pm$ &  2.77$\pm$ & -0.042$\pm$ & 6.54$\pm$&  448 &  452.5 &  1.01 \\
  & 7.4 & 0.009&  0.123&  0.632 & & &  \\
\hline

\end{tabular}
\end{center}
\caption{ BGK fit with fixed valence quarks for $\sigma_r$  for H1ZEUS-NC data in the range $Q^2 \ge 3.5$ or 8.5~GeV$^2$ and $x\le 0.01$. NLO fit.  { \it Soft gluon}.   $ m_{uds}= 0.14, m_{c}=1.3$ GeV.  $Q_0^2=1.9$ GeV$^2$.}
\label{tabl1}
\end{table}
\begin{table}[htbt]
\begin{center}
\begin{tabular}{|c|c|c|c|c|c|c|c|c|c|c|} 
\hline 
 $Q^2_{min}$ [GeV$^2$] &
$\sigma_0[mb]$ & $A_g$ & $\lambda_g$ & $C_g$ & $N_{df}$& $\chi^2$& $\chi^2/N_{df}$\\
\hline
 $3.5 $&  89.99$\pm$ &  2.44$\pm$ & -0.079$\pm$ & 7.24$\pm$&  530 &  540.35 &  1.02 \\
  & 9.2 & 0.145&  0.099&  0.61 & & &  \\
\hline
\end{tabular}
\end{center}
\caption{ BGK fit with fitted valence quarks for $\sigma_r$  for H1ZEUS-NC data in the range $Q^2 \ge 3.5$~GeV$^2$  and $x\le 0.01$. NLO fit. { \it Soft gluon}.   $ m_{uds}= 0.14, m_{c}=1.3$ GeV.  $Q_0^2=1.9$ GeV$^2$.}
\label{tabl2}
\end{table}

\begin{table}[htbt]
\begin{center}
\begin{tabular}{|c||c||c||c|c||c|c|c||c|c|c||c|} 
\hline 
No& 
$ Auv(fix)$& $Buv$ & $Cuv$&  $Euv$ & $Adv(fix)$& $Bdv$&
$Cdv$\\
\hline
1 &
4.073(sum rule) &  0.892$\pm$ &5.832$\pm$& 17.997$\pm$ &  3.151(sum rule) & 0.840$\pm$ & 3.480$\pm$\\
 &  & 0.019 & 0.341&  0.876& & 0.012& 0.056 \\
\hline
\end{tabular}
\end{center}
\caption{Parameters of the valence quark contribution fitted in the  BGK fit of Table~\ref{tabl2}.}
\label{tabl3}
\end{table}

\begin{table}[htbt]
\begin{center}
\begin{tabular}{|c|c|c|c|c|c|c|c|c|c|c|c|} 
\hline 
 $Q^2_{min}$ [GeV$^2$] &  $\sigma_0 [mb] $ &
 $A_g$ & $\lambda_g$ & $C_g$ &$D_g$ & $E_g$& $N_{df}$& $\chi^2$& $\chi^2/N_{df}$\\
\hline
  $3.5 $& 77,6$\pm$ &2.62$\pm$ & -0.064$\pm$&37.1$\pm$&3.06 $\pm$ & 1406.4$\pm$ & 532 & 534.2 & 1.00 \\
 & 18,6 & 0.16&  0.0087& 5.06 & 6.51 &  552.7 & & & \\
\hline
 8.5  & 63.5 $\pm$ & 2.11$\pm$ & -0.054$\pm$&21.3$\pm$&1.10 $\pm$ & 867.2$\pm$ & 448 & 439.0 &0.98 \\
 & 18.5 & 0.10&  0.0065& 4.062 & 5.76 &  423.7 &  & &\\
\hline
\end{tabular}
\end{center}
\caption{ BGK fit with valence quarks for $\sigma_r$  for H1ZEUS-NC data in the range $Q^2 \ge 3.5$ or 8.5~GeV$^2$ and $x\le 0.01$. NLO fit. { \it Soft + hard gluon}.   $ m_{uds}= 0.14, m_{c}=1.3$ GeV. $Q_0^2=1.9$ GeV$^2$.}
\label{tabl7}
\end{table}
\begin{table}[ht]
\begin{center}
\begin{tabular}{|c|c|c|c|c|c|c|c|c|c|c|c|} 
\hline 
$Q^2_{min}$ [GeV$^2$] & $Q^2_{0}$ [GeV$^2$] & 
 $\sigma_0 [mb] $ &
 $A_g$ & $\lambda_g$ & $C_g$ &$D_g$ & $E_g$& $N_{df}$& $\chi^2$& $\chi^2/N_{df}$\\
\hline
  $3.5 $ & 1.1 & 220 $\pm$ &3.57$\pm$ & 0.082$\pm$&31.3$\pm$& 11.0 $\pm$ & 1360$\pm$ & 532 & 532 & 1.00 \\
 &  & 122 & 0.38&  0.017& 5.4 & 9.8 &  690 & & & \\
\hline
\end{tabular}
\end{center}
\caption{ BGK fit with valence quarks for $\sigma_r$  for H1ZEUS-NC data in the range $Q^2 \ge 3.5$~GeV$^2$ and $x\le 0.01$. NLO fit.  { \it Soft + hard gluon}.   $ m_{uds}= 0.14, m_{c}=1.3$ GeV. $Q_0^2=1.1$ GeV$^2$.}
\label{tabl9}
\end{table}

In our previous work~\cite{kowlusz}, which used a subset of data from the present evaluation and had a much poorer evaluation of errors, the differences between the fits with and without valence quarks were quite pronounced. Therefore, in Table~\ref{tabl6} we show a BGK fit without the valence quarks. We observe that the fit is only slightly worse than the one with the fixed valence quarks, see Table~\ref{tabl1} and~\ref{tabl2}.  
The correlated errors of the final data are  more restrictive than the uncorrelated one, as seen from the standard HERAPDF fits, however, in the case of dipole fits, the full treatment of errors  makes the contribution of valence quarks less visible in the low $x$ region.

\begin{figure}[htbp]
\centering
\includegraphics[width=12.0cm]{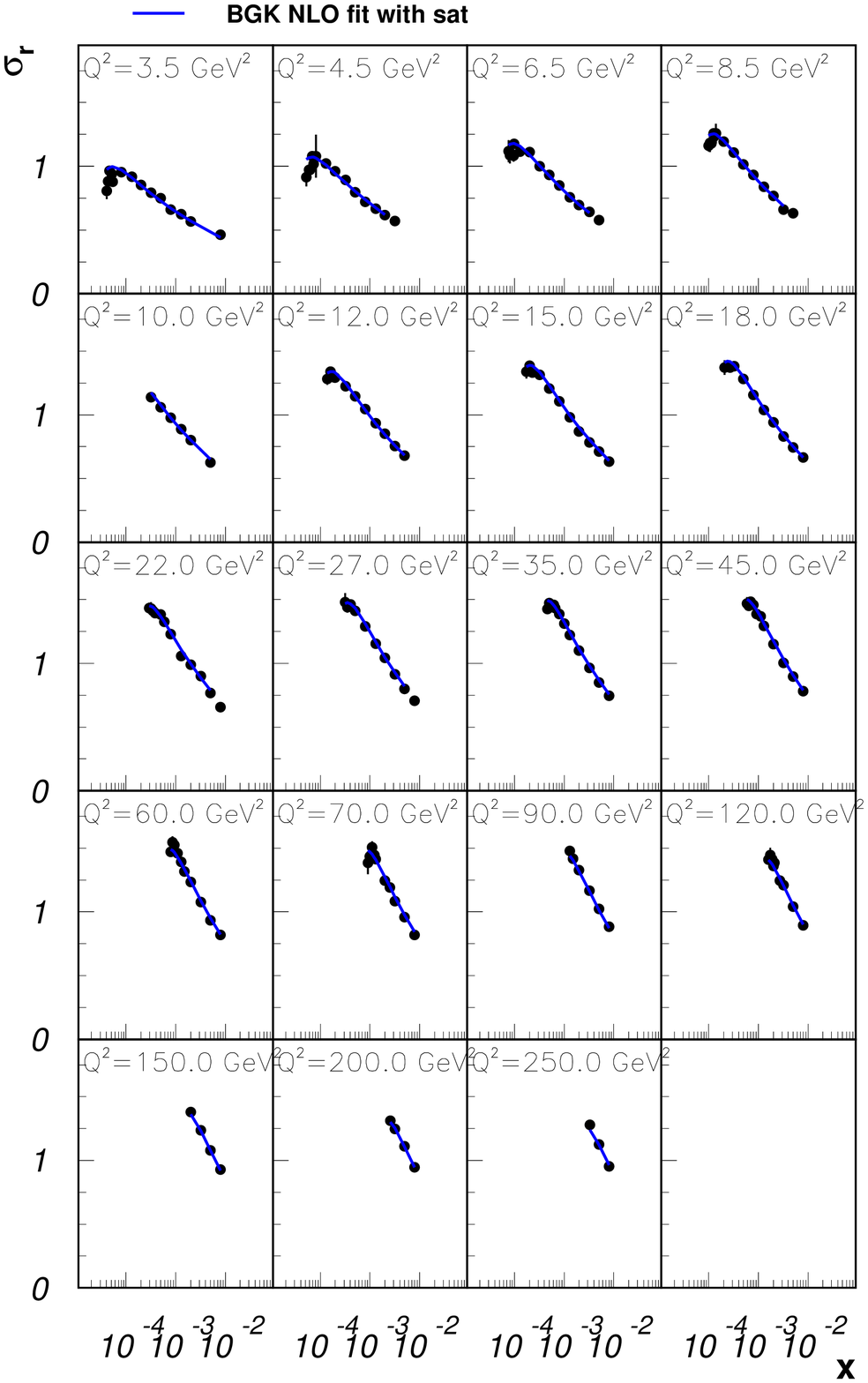}
\caption{Comparison  of the dipole BGK fit of Table~\ref{tabl7} with the reduced cross sections  of the final, combined H1 and ZEUS  HERA  data. For better visibility only 920 GeV data are displayed. The fit was performed in the $Q^2 > 3.5$ GeV$^2$ region.}
\label{bgk-fit}
\end{figure}

In Fig.~\ref{bgk-fit} we show a comparison of the dipole BGK fit  with valence quarks and soft+hard gluon density, Table~\ref{tabl7},  with the data of HERA at $Q^2 > 3.5$ GeV$^2$. For clarity only  reduced cross section data taken with $E_p = 920$ GeV is shown. Figure shows an excellent agreement  with data. 
\begin{table}[ht]
\begin{center}
\begin{tabular}{|c|c|c|c|c|c|c|c|c|c|c|c|} 
\hline 
$Q^2_{min}$ [GeV$^2$]&
$\sigma_0 [mb]$ & $A_g$ & $\lambda_g$ & $C_g$ & $N_{df}$& $\chi^2$& $\chi^2/N_{df}$\\
\hline
 3.5 & 105.20$\pm$ &  2.4788$\pm$ & -0.066$\pm$3 &  6.9093 $\pm$&  534 & 554.68  & 1.04 \\
  & 12.234 & 0.093&  0.004&  0.510 & & &  \\
\hline
\end{tabular}
\end{center}
\caption{ BGK fit without valence quarks for $\sigma_r$  for H1ZEUS-NC data in the range $Q^2 \ge 3.5$~GeV$^2$ and $x\le 0.01$. NLO fit. { \it Soft gluon}.   $ m_{uds}= 0.14, m_{c}=1.3$ GeV. $Q_0^2=1.9$ GeV$^2$.}
\label{tabl6}
\end{table}

In this investigation the values of the light quark mass in the dipole formula were fixed to 0.14 GeV, like in the original GBW and BGK models. We performed also fits lowering the light quark masses but the resulting fits were of similar quality as the fits shown here.

\subsection{Results from the pdf fits}
In the Table~\ref{tabl4} we show results of the standard HERAPDF fits~\cite{hiH1ZEUS} to the final HERA inclusive cross section data, which have  fully correlated errors. They  are performed in the same Q$^2$ ranges as the dipole fits but in the full $x$ range. The full $x$ range is here necessary  to fix the contribution of  valence quarks. In Table~\ref{tabl5} we show the parameters of the valence quarks obtained for $Q^2 > 3.5$ GeV$^2$. These are the parameters used for the dipole fit with fixed valence quarks shown in Table~\ref{tabl1} and~\ref{tabl7} .   

\begin{table}[ht]
\begin{center}
\begin{tabular}{|c|c|c|c|c|c|c|c|c|c|c|c|} 
\hline 
No& 
$Q^2_{min}$ [GeV$^2$] &HF Scheme&
$Np$& $\chi^2$& $\chi^2/Np$\\
\hline
1 &
 3.5  & RT& 1131 &  1356.70 & 1.20 \\
\hline
2 &
 8.5 & RT&  456 &  470.88 & 1.15 \\
\hline
\end{tabular}
\end{center}
\caption{HERAPDF NLO fits to the same data set as for the dipole model but in the full $x$ range. $Q_0^2=1.9$ GeV$^2$.}
\label{tabl4}
\end{table}
\begin{table}[h]
\begin{center}
\begin{tabular}{|c||c||c||c|c||c|c|c||c|c|c||c|} 
\hline 
No& 
$ Auv$& $Buv$ & $Cuv$&  $Euv$ & $Adv$& $Bdv$&
$Cdv$\\
\hline
1 &
  4.073$\pm$ &0.713$\pm$ &4.841$\pm$& 13.405$\pm$ &  3.151$\pm$1&0.806$\pm$ &4.079$\pm$\\
  &0.123  & 0.016 & 0.214&  0.921& 0.121& 0.056& 0.301 \\ 
\hline
\end{tabular}
\end{center}
\caption{Parameters of valence quarks obtained in HERAPDF NLO fits for  $Q^2 > 3.5$ GeV$^2$.}
\label{tabl5}
\end{table}
\begin{table}[htbt]
\begin{center}
\begin{tabular}{|c|c|c|c|c|c|c|c|c|c|c|c|} 
\hline 
No& 
$Q^2_{min}$ [GeV$^2$] &HF Scheme&
$Np$& $\chi^2$& $\chi^2/Np$\\
\hline
 1 &
3.5  &FONLL-B& 534 &   539,3 & 1.01 \\
2 &
 3.5  & FONLL-B& 532 &   537,3 & 1.01 \\
\hline
\end{tabular}
\end{center}
\caption{HERAPDF NLO fits with fixed valence quarks to the same data set as for the dipole model, but with $x< 0.01$ range. $Q_0^2=1.9$ GeV$^2$. 
No 1 {\it soft gluon}, No 2{ \it soft + hard gluon}}
\label{tabl4a}
\end{table}


Table~\ref{tabl4} shows that the standard HERAPDF fit is not describing data very well. The agreement improves somewhat when the fit is performed in a higher $Q^2$ range but it is still not fully satisfactory, as was extensively discussed in   ref.~\cite{hiH1ZEUS}. We note that in case of the BGK dipole model the agreement with data is very good, see Table~\ref{tabl1} and~\ref{tabl2}. It is even slightly improving when a 5 parameter ansatz for gluon density, {\it soft + hard}, is used, see Table~\ref{tabl7} and~\ref{tabl9}.   The quality of a fit is not depending on the starting scale, an example of a fit with $Q_0^2=1.1$ GeV$^2$ is shown in Table~\ref{tabl9}.  
In difference to the results of our previous paper~\cite{kowlusz}, the dipole fit quality is not significantly improving with increasing $Q^2_{min}$, see Tables~\ref{tabl1} and~\ref{tabl7}.  

It is also interesting to observe that in the low $x$ region, the dipole and HERAPDF fit have similar quality, see Table~\ref{tabl7} and~\ref{tabl4a}. The HERAPDF fits  of  Table~\ref{tabl4a} were performed with the fixed quark contribution and with the same ansatz for gluon density,  {\it soft gluon} and { \it soft + hard gluon}, as in the dipole case. 
However, the results of the HERAPDF fit in the low $x$ region only are sizably scheme dependent. In the standard HERAPDF RT-OPT scheme~\cite{rt.opt} the fit quality is somewhat poorer, $\chi^2/N_{df}=1.06$, instead of  $\chi^2/N_{df}=1.01$ in  FONLL-B~\cite{fonellb}.

\begin{figure}[htbt]
\centering
\includegraphics[width=9.0cm]{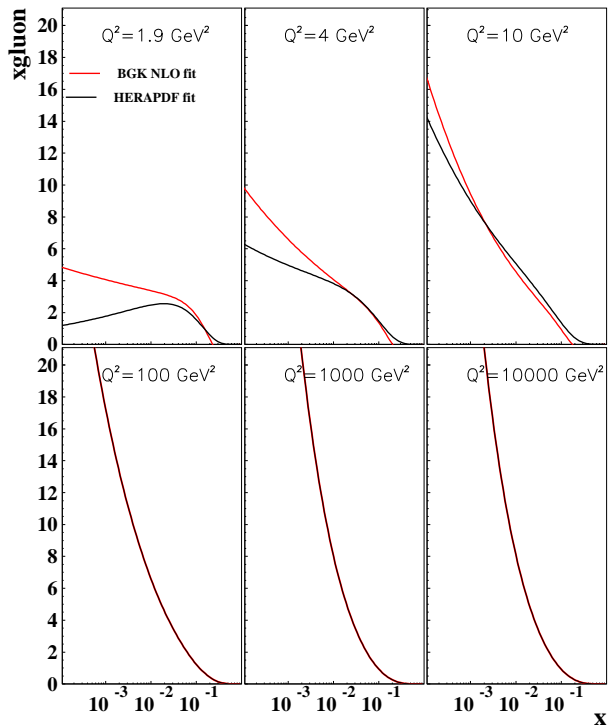}
\caption{Comparison between the dipole (soft), Table \ref{tabl1},  and HERAPDF gluon (soft) in NLO. }
\label{glu-nlo}
\end{figure}

In Fig.~\ref{glu-nlo} we show a comparison of the gluon density obtained in the fits with valence quarks and compare it to the gluon density obtained in the HERAPDF fit.  We see that the two gluon densities differ, even substantially, at smaller scales but then start to closely approach each other at higher scales.    

{\bf Summarizing} we can tell that  the dipole BGK fits describe the highest precision HERA data, in the  low $x$ region and for $Q^2>3.5$ GeV$^2$, very well. The best fits were obtained with the five parameter form of the gluon density and with the saturation ansatz, Table~\ref{tabl7} and Figure~\ref{bgk-fit}. The differences in the fit quality, between various  fits performed in this region are however pretty small. In our previous work~\cite{kowlusz}, which used a subset of data from the present evaluation and had a much poorer  evaluation of errors, the differences between the fits with and without valence quarks were more pronounced. The correlated errors of the final data are  more restrictive than the uncorrelated one, as seen from the standard HERAPDF fits, however, in the case of dipole fits, the full treatment of errors  makes the contribution of valence quarks less visible in the low $x$ region. It is also interesting to observe that in difference to the previous fits (with  uncorrelated errors)~\cite{kowlusz}, the present fits   have a similar quality in the higher and  lower $Q^2$ region. A substantial improvement of the fit quality with the increase of $Q^2$ cut, observed in the previous evaluation, like $\chi^2/N\approx 1$ for $Q^2>3.5$ GeV$^2$ and $\chi^2/N\approx 0.8$ for $Q^2>8.5$ GeV$^2$~\cite{kowlusz}, seemed to suggest some kind of saturation or lack of higher order QCD corrections. This is now not seen anymore, all fits seem to be of similar quality.

\section{Investigation of saturation effects}

Saturation is a property of a gluonic state, which is so dense  that gluons start to interact with each other.  
In the BGK dipole model, saturation effects are described in the  eikonal approximation,  eq.~\ref{eBGK}. To investigate the importance of saturation at HERA  we performed also fits without the saturation ansatz,  i.e when only the first term in the expansion of the exponent of eq.~\ref{eBGK} is taken into account, i.e.
\begin{equation}
\label{eBGKns}
   \sigma_{\text{dip}}(x,r^{2}) = \pi^{2} r^{2} \alpha_{s}(\mu^{2}) xg(x,\mu^{2})/3.
\end{equation}
  In  Table~\ref{tabl8}  we show the results of such a fit at two starting scales, $Q_0^2 =1.9$ and 1.1 GeV$^2$. Both fits describe the data fairly well although  the fit quality is slightly worse than in the fits with the saturation ansatz, Table~\ref{tabl7} and~\ref{tabl9}. 
\begin{table}[ht]
\begin{center}
\begin{tabular}{|c|c|c|c|c|c|c|c|c|c|c|c|} 
\hline 
$Q^2_{min}$ [GeV$^2$] &  $Q^2_{0}$ [GeV$^2$] &
 $A_g$ & $\lambda_g$ & $C_g$ &$D_g$ & $E_g$& $Ndf$& $\chi^2$& $\chi^2/Np$\\
\hline
  3.5  & 1.9 & 2.33$\pm$ & -0.094$\pm$&14.8$\pm$&9.80 $\pm$ & -99.5$\pm$ & 533 &  556.17  & 1.04 \\
  & & 0.10&  0.006& 11.5 &14.7 &  74.830 & & & \\
\hline
  3.5  & 1.1 &3.80$\pm$ & 0.10$\pm$&32.5$\pm$&-25.2 $\pm$ & 1868 $\pm$ & 533 &  539.2  & 1.01 \\
 &  & 0.22&  0.01& 1.6 &3.49 &  252 & & & \\
\hline  
\end{tabular}
\end{center}
\caption{ BGK fit with valence quarks for $\sigma_r$  for H1ZEUS-NC data in the range $Q^2 \ge 3.5$~GeV$^2$ and $x\le 0.01$. NLO fit.  { \it Soft + hard gluon}.   $ m_{uds}= 0.14, m_{c}=1.3$ GeV, non-saturation ansatz. $Q_0^2=1.9$ or 1.1 GeV$^2$.}
\label{tabl8}
\end{table}

We observe that the values of the parameter $\sigma_0$ of dipole cross section are quite high, of the order 70 mb for $Q^2_0 = 1.9$ GeV$^2$ and 220 mb for $Q^2_0 = 1.1$ GeV$^2$, see Table~\ref{tabl7} and~\ref{tabl9} . This is much higher than in the original GBW and BGK model fits~\cite{GBW,BGK}, where this number was around 23 mb. This is an interesting result because $\sigma_0$  is the black disk limit of the  dipole cross section, i.e. its value at very large energies.   It indicates that the exponential form of the dipole cross section may be of little importance because, in the limit of very high values of $\sigma_0$, the dipole cross section reduces to the first term of the expansion of the exponent in the dipole cross section, see eq~\ref{eBGKns}. This is in agreement with the fit results performed with and without saturation shown in Table~\ref{tabl7},~\ref{tabl9} and~\ref{tabl8}. 

\begin{figure}[htbt]
\centering
\includegraphics[width=9.0cm]{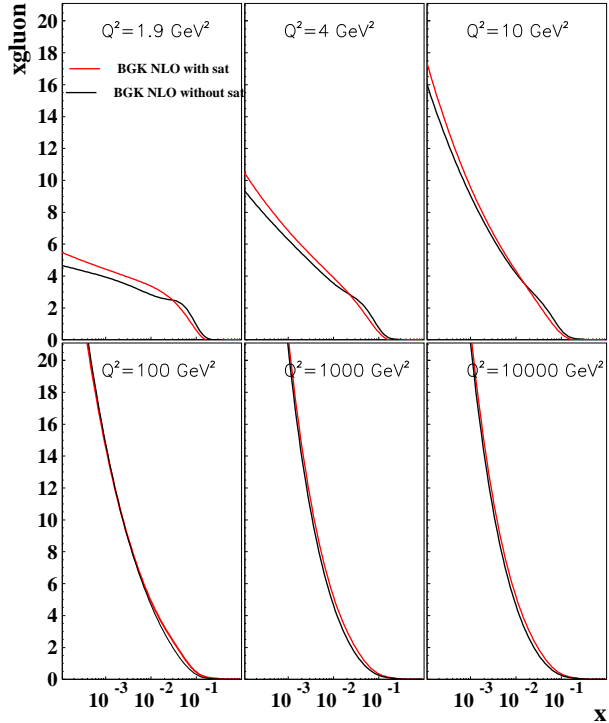}
\caption{Comparison between the gluon densities obtained from fits in the $Q^2 > 3.5 $ GeV$^2$ region, with and without the saturation ansatz, Table~\ref{tabl7} and \ref{tabl8}. }
\label{glu-sns}
\end{figure}

In Figure~\ref{glu-sns} we compare the gluon densities obtained from fits with and without saturation ansatz of Tables~\ref{tabl7} and~\ref{tabl8}.
Note that for $x< 0.001$, the gluon densities obtained in the dipole approach are higher than that of the standard pdf fit, Fig.~\ref{glu-nlo}, and that of the non-saturated one, see Fig~\ref{glu-sns}. This is expected  and is due to damping  of gluon density by  saturation effects.

The fits of Table~\ref{tabl7},~\ref{tabl9} and~\ref{tabl8} where performed in the higher $Q^2> 3.5$ GeV$^2$ region. Although they show some slight worsening of the fit quality between the saturated and non-saturated case, the 
 differences are by themselves too small to be considered as an indication of saturation. The situation changes however when we start to look in the region of smaller  $Q^2$'s, less than 3.5 GeV$^2$.

The degree of saturation, in DIS,
is characterized by the size of the dipole, $r_S$, which, at a given $x$,  starts to interact multiple times in a proton (in about 60\% of cases). It is usually expressed as a saturation scale  $Q_S^2=2/r_S^2$, which  in the GBW model is given by  $Q_S^2=2/R_0^2$. In the BGK (or KMW) model it can be determined directly, from the gluon density, using eq.~\ref{eBGKns} (or its analog in KMW). In the complete, impact parameter dependent analysis of~\cite{KMW}, it 
  was determined at HERA as $Q_S^2=0.5$ GeV$^2$ at $x=10^{-3}$ and as about 1 GeV$^2$ at $x=10^{-4}$.  

\begin{figure}[htbp]
\centering
\includegraphics[width=12.0cm] {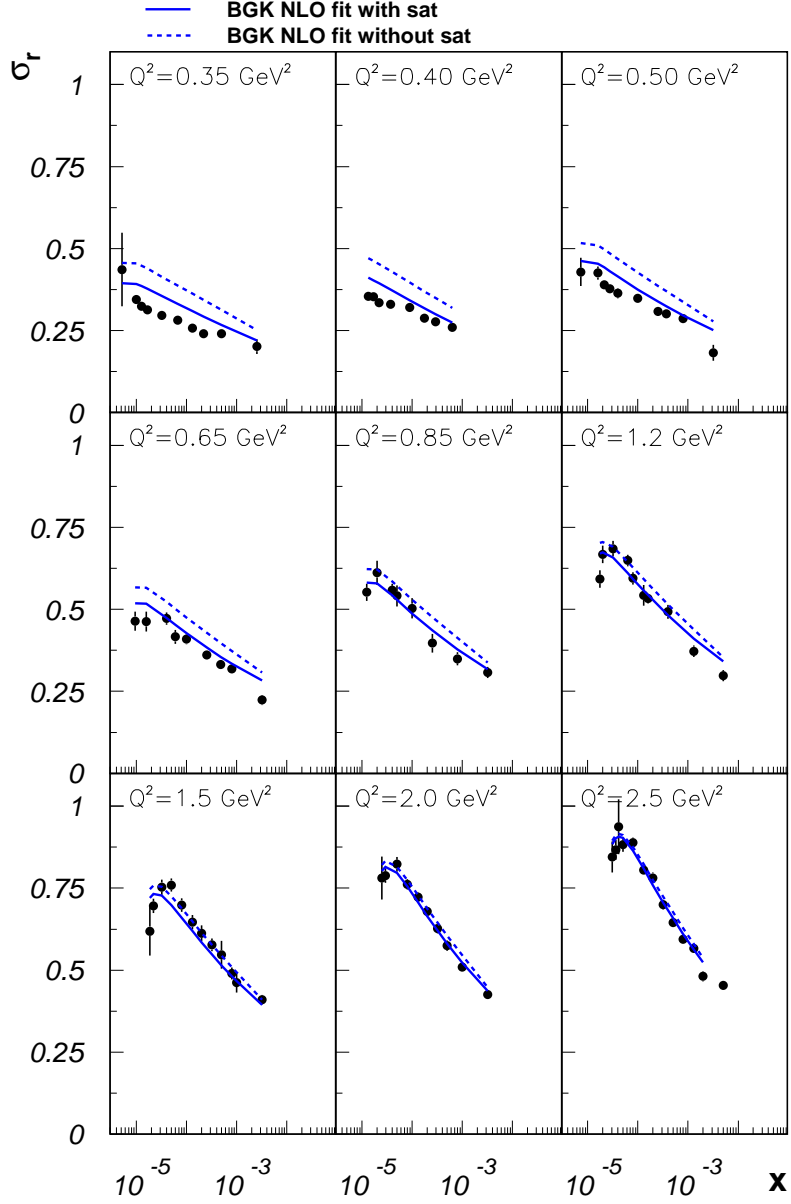}
\caption{  Comparison  of the {\it extrapolated} dipole BGK fit of Table~\ref{tabl7} and~\ref{tabl8}   with the reduced cross sections  of the final, combined H1 and ZEUS  HERA  data. For better visibility only 920~GeV data are displayed.  The starting scale of the evolution was $Q^2_0 = 1.9$ GeV$^2$.  The fit was performed in the $Q^2 > 3.5$ GeV$^2$ region and then extrapolated to the lower $Q^2$ region seen in the figure. Solid line shows the extrapolation of the fit with saturation and the dashed one without saturation.}
\label{bgk-ext}
\end{figure}

The low value of the saturation scale, $Q^2_S$, determined in~\cite{KMW}, suggests that we should be looking for saturation effects in the low $Q^2$ data region.
 In  Fig.~\ref{bgk-ext}, we show,  therefore, the results of the extrapolation  to a low $Q^2$ region, $Q^2 < 3.5$ GeV$^2$,  of the fits with and without saturation of Tables~\ref{tabl7} and~\ref{tabl8}.  We see that the fit with saturation of Table~\ref{tabl7} (solid line) extrapolates down to  $Q^2 =  0.85$~GeV$^2$ fairly well. The extrapolation to even lower $Q^2$  starts to overshoot the data in a systematic way. 
 The fit without saturation, of Table~\ref{tabl8} (dashed line), extrapolates  well  to  $Q^2 =  1.2$~GeV$^2$ only. The extrapolation to the  lower $Q^2$ region overshoots the data sizably stronger than in the fit with saturation. 
 The $\chi^2/N_{df}$ of the fit  which uses the saturation ansatz (solid line), including the extrapolated points,  is 1.24. For the fit  without saturation (dashed line) it is 1.6. 
 
 We also fitted data with and without the saturation ansatz in the whole $Q^2$ region, $0.35 < Q^2< 250$ GeV$^2$, and found that the fit is only slightly better than in the extrapolated case. The results are given in Table~\ref{tabl11} for the saturated case and in Table~\ref{tabl12} for the non-saturated fit.  
 \begin{table}[ht]
\begin{center}
\begin{tabular}{|c|c|c|c|c|c|c|c|c|c|c|c|} 
\hline 
  $Q^2_{0}$ [GeV$^2$] & $\sigma_0$ [mb] &
 $A_g$ & $\lambda_g$ & $C_g$ &$D_g$ & $E_g$& $N_{df}$& $\chi^2$& $\chi^2/N_{df}$\\
\hline
$1.9 $ & 38.2 $\pm$   & 2.80 $\pm$ & -0.063  $\pm$ & 46.3$\pm$& 12.1 $\pm$ & 1970.4 $\pm$ &653 &790.4& 1.21 \\
  &4.1 &0.14& 0.006 & 4.58 &6.00 & 566.0 & &  &\\  
\hline
 $1.1 $&  196,1 $\pm$ &6.24 $\pm$ &  0.098 $\pm$ &52.3 $\pm$& -22.0 $\pm$ & 2145.0  $\pm$ &653& 894.1 & 1.37 \\
  &105 & 0.53& 0.012& 6.5 & 10.64 &  835.7 & &  &\\
\hline  
\end{tabular}
\end{center}
\caption{ BGK fit with valence quarks for $\sigma_r$  for H1ZEUS-NC data in the range $Q^2 \ge 0.35$~GeV$^2$ and $x\le 0.01$. NLO fit.  { \it Soft + hard gluon}.   $ m_{uds}= 0.14, m_{c}=1.3$ GeV, saturation ansatz. $Q_0^2=1.9$ or 1.1 GeV$^2$.}
\label{tabl11}
\end{table}
 \begin{table}[ht]
\begin{center}
\begin{tabular}{|c|c|c|c|c|c|c|c|c|c|c|c|} 
\hline 
  $Q^2_{0}$ [GeV$^2$] &
 $A_g$ & $\lambda_g$ & $C_g$ &$D_g$ & $E_g$& $N_{df}$& $\chi^2$& $\chi^2/N_{df}$\\
\hline  
$1.9 $&  3.05  $\pm$ &  -0.022  $\pm$ & 40.3 $\pm$&-32.3 $\pm$ &  3158.3 $\pm$ &654&1024.3& 1.56 \\
  &0.092& 0.004 & 1.067 &3.02 & 219.3 & & \\
\hline
$1.1 $&5.62 $\pm$ & 0.158  $\pm$ & 43.320  $\pm$& -55.011  $\pm$ & 3791.6$\pm$ &654& 999.98 &1.53 \\
 &0.13& 0.001 &  0.15 & 8.62 & 187.7 & & \\
\hline  
\end{tabular}
\end{center}
\caption{ BGK fit with valence quarks for $\sigma_r$  for H1ZEUS-NC data in the range $Q^2 \ge 0.35$~GeV$^2$ and $x\le 0.01$. NLO fit.  { \it Soft + hard gluon}.   $ m_{uds}= 0.14, m_{c}=1.3$ GeV, non-saturation ansatz. $Q_0^2=1.9$ or 1.1 GeV$^2$.}
\label{tabl12}
\end{table}
For both starting scales, the fits with the saturated ansatz are sizably closer to data, which indicates a presence of saturation effects. The best fit is obtained with the saturated fit of Table~\ref{tabl11}, at the QCD scale  $Q_0^2=1.9$, which has  $\chi^2/N_{df}=1.21$.  The value of the parameter  $\sigma_0$, which is a black disc limit of virtual photon-proton cross section, has the smallest error and is within 2 standard deviations consistent with the fit at higher $Q^2$, of Table~\ref{tabl7}.  Its value of around 40 mb is also close to the value of the same parameter obtained in previous fits~\cite{GBW,BGK}, which were around 20 mb.

In the saturation investigation of ref~\cite{BGK,KT}, were the first set of HERA data was used, the fits with and without saturation had the same quality, even when the low $Q^2 $ region was included in the fit. Therefore, the present result that the fit with saturation has a sizably better quality than that without saturation is new and is due to the substantially improved quality of HERA data.  
 
Finally let us also note, that the lack of very good description of data, expressed by the worsening of the $\chi^2/N_{df}$ value from 1.00  to 1.21 when the  lower $Q^2$ region is included, together with  the systematic overshoot of the fits over data observed in Fig.~\ref{bgk-ext} at $Q^2 <  1$~GeV$^2$, suggests that the approach to saturation realized in the BGK model may  be too crude. 
 The large discrepancies between the values of  $\sigma_0$ parameter of Tables~\ref{tabl11},~\ref{tabl7} and~\ref{tabl9} together with their large measurement errors suggest that the saturation investigation presented here should be extended using an impact parameter dependent dipole model~\cite{KMW,KT}. In such a model, the QCD evolution and the saturation ansatz is the same as in the BGK model, however,  the   $\sigma_0$ parameter is replaced by the transverse profile of the proton.  The proton profile, which determines the impact parameter distribution, is obtained from the data of the exclusive diffractive scattering of vector mesons. Therefore, there is effectively  one free parameter less,  which could lead to an improved  investigation of the saturation mechanism. We intend to come back to this subject in an extended evaluation of HERA data which will also include the final, exclusive diffractive scattering HERA data.      
        The value of the investigation performed in this paper lays in its simplicity; in spite of the fact that the BGK model has one free parameter more than the impact parameter dependent model, it provides a clear evidence that saturated dipole mechanism is closer to data than the non-saturated one.

\section{Summary}
We found that the dipole BGK fits with the DGLAP QCD evolution  describe the highest precision HERA data, in the  low $x$ region and for $3.5 < Q^2>250$ GeV$^2$, very well. The best fits were obtained with the five parameter form of the gluon density and with the saturation ansatz, Table~\ref{tabl7} and Figure~\ref{bgk-fit}. The differences in the fit quality, between various  fits performed in this region are however pretty small. 

The present paper is focused on the saturation question, which is investigated by various fits to data with and without  the saturation ansatz. The fits were made with different gluon densities and using different QCD starting scales. All dipole fits to data in the higher $Q^2 > 3.5$ GeV$^2$ region, are  describing data very well. No significant differences between the saturated and no-saturated fits were observed. The fits to data including the lower  $0.35 < Q^2 < 3.5$ GeV$^2 $ region, show however, that the saturated gluon density is preferred. This is a new result, which is due to a substantial improvement of data quality obtained in the final evaluation of HERA data. It indicates also that there is more information about saturation in  HERA data, which should be evaluated using impact parameter  dependent dipole models. 
  

\section{Acknowledgement}
We would like to thank Sasha Glazov (DESY) for various help and useful discussions from the beginnig of the analysis presented here till the final version of the manuscript. We also thank xFitter developers' group for useful comments and Pavel Belov for careful reading of the manuscript. This work is supported by the Polish Ministry under
program Mobility Plus, no. 1320/MOB/IV/2015/0.

\end{document}